# Determining effective permeability at reservoir scale: Numerical simulations and theoretical modeling

Barnabas Adeyemi[1], Behzad Ghanbarian[1*], C. L. Winter[2], Peter R. King[3]

[1] Porous Media Research Lab, Department of Geology, Kansas State University, Manhattan KS 66506, USA

[2] Department of Hydrology and Atmospheric Sciences, University of Arizona, USA

[3] Department of Earth Science and Engineering, Imperial College London, London, UK

[*] Corresponding author's email address: ghanbarian@ksu.edu

## Abstract

Determining the effective permeability ($k_{\text{eff}}$) of geological formations has broad applications to site remediation, aquifer discharge or recharge, hydrocarbon production, and enhanced oil recovery. The objectives of this study are: (1) to explore an approach to estimating $k_{\text{eff}}$ at the reservoir scale using the critical path analysis (CPA), (2) to evaluate the accuracy of this new approach by comparing the estimated $k_{\text{eff}}$ to the numerically simulated effective permeability, and (3) to compare the performance of CPA estimates of $k_{\text{eff}}$ to estimates by three other models i.e., perturbation theory (PT), effective-medium approximation (EMA), and renormalization group theory (RGT). We construct two- and three-dimensional random (uncorrelated) geologic formations based on permeability measurements from the Borden site and assume that the permeability distribution conforms to the log-normal probability density function over a wide range of means and standard deviations. Comparing $k_{\text{eff}}$ estimated via CPA to $k_{\text{eff}}$ values derived from numerical flow simulations indicates that CPA provides accurate estimations in both two and three dimensions over a wide range of heterogeneity levels, similar to RGT. Inter-model comparisons show that although PT and EMA provide reasonable $k_{\text{eff}}$ estimations in rather homogeneous formations, they substantially overestimate the effective permeability in highly heterogeneous formations.

## 1. Introduction

Investigating flow and transport in geological formations, such as aquifers and reservoirs, is essential in numerous areas of geology and engineering: for $CO_2$ sequestration, site remediation, groundwater hydrology, and enhanced oil recovery. Under fully saturated conditions, one key parameter is effective permeability ($k_{\text{eff}}$), which indicates the overall capability of a formation to allow the passage of fluid through it. Geological formations are heterogeneous and typically composed of zones of various materials with different permeabilities spanning several orders of magnitude (Freeze and Cherry, 1979; Akpoji and De Smedt, 1993; Oladele et al., 2019). Due to the presence of heterogeneity across scales, accurate calculation of $k_{\text{eff}}$ requires precise characterization of reservoirs. However, detailed field observations of local reservoir characteristics are typically not available, which puts a premium on analysis based upon realistic theoretical and computational models.

Experimental studies (Bjerg et al., 1992; Rehfeld et al., 1992; Sudicky, 1986) show that the histogram of permeability values measured on core samples approximately follows the log-normal probability density function. In fact, the log-normal permeability distribution has been widely used to study flow and transport at large scales (Colecchio et al., 2020; Edery et al., 2014; Hristopulos, 2003; Zarlenga et al., 2018). This implies that the spatial heterogeneity of a formation can be captured by truncated log-normal distribution parameters, i.e., mean and standard deviation as well as its lower and upper cutoffs.

An active subject of research in geosciences has been determining the effective value of permeability at the continuum scale (Dagan, 1993; Masihi et al., 2016; Rasaei and Sahimi, 2009). Various techniques, including theoretical (Renard and de Marsily, 1997; Sanchez-Vila et al., 2006), numerical (Akai et al., 2019; Vu et al., 2018), and machine learning-based (Da Wang et al., 2021; Mudunuru et al., 2020) methods, have been proposed to calculate the value of $k_{\text{eff}}$ in geological formations. Although numerical methods are suitable for any type of aquifers and reservoirs, they are computationally demanding, particularly for three-dimensional (3D) reservoirs. As a result, theoretical models and more recently machine learning approaches have been frequently utilized for the estimation of $k_{\text{eff}}$. Theoretical models include the simple averaging techniques (Deutsch, 1989), perturbation theory (King, 1987), self-consistent approximation (Dagan, 1979), effective medium approximation (Fokker, 2001), renormalization group theory (King, 1989), wavelet transformation (Rasaei and Sahimi, 2009), and information theory (Wood and Taghizadeh, 2020). Because many models in the hydrology literature are based on perturbative methods (Sanchez-Vila et al., 2006), we briefly review them in what follows.

The organization of the paper is as follows. We first describe concepts of perturbation theory, critical path analysis, renormalization group theory, and effective-medium approximation. In Section 3, objectives of this study are explained. In Section 4, generation of geologic formations and numerical simulations of flow are described. Section 4 presents the results, and Section 5 discusses model performances.

## 2. Theoretical upscaling approaches

### 2.1. Perturbation theory

Within the framework of PT, the pressure head in the Darcy equation is first expanded in a power series in terms of permeability fluctuations (Sanchez-Vila et al., 2006) after which a solution for velocity is constructed by applying Darcy's law (Sanchez-Vila et al., 2006; Renard and de Marsily, 1996; Stepanyants and Teodorovich, 2003). Using these methods, Matheron (1967) and Gutjahr (1978) proved that the quantities of the equation popularly known in hydrogeology as the Matheron's conjecture are the first two terms of the Taylor series expansion of an exponential function. Although not generally proven to be exact in 3D flow, the conjecture is known to give the effective permeability in log-normally distributed permeability fields as the harmonic mean ($k_{eff} = k_h$) in one-dimensional (1D) flow and geometric mean ($k_{eff} = k_g$) in two-dimensional (2D) flow (De Wit, 1995).

Indelman and Abramovich (1994) showed that for an anisotropic permeability field $k_{eff}$ depends not only on the anisotropic ratios, variance, and space dimensions but also on the shape of the permeability distribution function. Importantly, their work highlighted major inconsistencies in Matheron's conjecture for anisotropic and 3D systems. Their expression, which we denote here as the anisotropic perturbation theory (ANPT), is

$$k_{\text{eff}} = k_g \left\{ 1 + \left(\frac{1}{2} - \alpha_i\right)\sigma_Y^2 + \frac{1}{2}\left[\left(\frac{1}{2} - \alpha_i\right)^2 + \gamma_i\right]\sigma_Y^4 \right\} \tag{1}$$

where $\sigma_Y$ is the standard deviation of the natural logarithm of the permeability (Y = ln(k)), $i$ indicates the principal hydraulic conductivity direction ($i$ = 1,2,3 in three dimensions), and $\alpha_1 = \alpha_2 = (1 - \chi)/2$ and $\alpha_3 = \chi$. $\chi$ depends on the anisotropic ratio of the permeability field and $\gamma_i$ depends on the permeability correlation function. In the case of isotropy, $\chi$ = 1/3 and $\gamma_i = 0$.

De Wit (1995) derived another expression for the $k_{\text{eff}}$ that exposed some of the underlying inaccuracies of Matheron's conjecture as the $\sigma_Y^6$ order terms in his derivation contained parameters and factors that are not available in the $\sigma_Y^6$ order terms of the Matheron's conjecture expansion. The expression, here referred to as the simple perturbation theory (SPT), is:

$$k_{\text{eff}} = k_g \left[ 1 + \left(\frac{1}{2} - \frac{1}{d}\right)\sigma_Y^2 + \frac{1}{2}\left(\frac{1}{2} - \frac{1}{d}\right)^2 \frac{\sigma_Y^4}{2} + \left(\frac{1}{2} - \frac{1}{d}\right)^3 \frac{\sigma_Y^6}{2} + \varepsilon \right], \quad \varepsilon = -\frac{\sigma_Y^6}{3n^3} + \frac{\beta}{n} \tag{2}$$

where $d$ is the system dimensionality ($d$ = 3 in three dimensions), and $\varepsilon$ is a term that depends on the permeability distribution function and vanishes for $d$ = 1 and 2. For three-dimensional flow, however, it was numerically found that $\varepsilon$ is approximately equal to $-0.0014\sigma_Y^6$ for a Gaussian log permeability field (De Wit, 1995; Sanchez-Vila, 2006).

More recently, Stepanyants and Teodorovich (2003) used a different perturbative approach to construct a perturbation series and calculate the effective permeability. Their approach led to a solution presented in the form of a power series for the inverse coefficient of permeability, which we refer to as the alternative perturbation theory (ALPT), is given by

$$k_{\text{eff}} = k_g \exp\left(-\frac{\sigma_Y^2}{2}\right)\left[1 - \frac{d-1}{d}\sigma_Y^2 + \frac{1}{2}\left(\frac{d-1}{d}\right)^2 \sigma_Y^4\right]^{-1} \tag{3}$$

Although the derivation of most perturbative models requires advanced mathematical and computational skills, the inability of these methods to accurately estimate $k_{\text{eff}}$ in heterogeneous formations where permeability fluctuations become very large (King, 1989; Sanchez-Vila et al., 2006; Dagan et al., 2012) is well known in the literature, as we also show in this study. Such results demonstrate the inability of perturbative methods to accurately estimate the effective permeability in statistically heterogeneous formations.

**2.2. Critical path analysis**

Critical path analysis (CPA) was originally proposed in the physics literature to scale up conductivity in random (uncorrelated) and heterogeneous systems with large fluctuations in local conductivity (Ambegaokar et al., 1971; Pollak, 1972). Based on the CPA, fluid flow in a heterogeneous formation with a broad distribution of permeabilities is controlled by permeabilities whose magnitudes are greater than some critical permeability (Hunt et al., 2014). In other words, transport is dominated by high-permeability zones, while low-permeability ones have trivial contribution to the overall transport (Černý, 2004; Hunt, 2001).

Imagine a reservoir constructed of grid blocks of various permeabilities. To calculate the value of critical permeability, one should first remove all the grid blocks from the reservoir. one should then replace them sequentially in their original locations in a decreasing order from the largest to the smallest permeability. As the first largest permeabilities are replaced, there is still no percolating cluster. However, after a sufficiently large fraction of grid blocks is replaced within the reservoir, a sample-spanning cluster forms and the system starts percolating. The critical permeability is defined as the smallest permeability required to form a conducting sample-spanning cluster. Fluid flow and transport take place through the sample-spanning cluster which is composed of two components: (1) the dead-end part that does not participate to flow, and (2) the backbone, the multiply-connected part of the cluster, through which fluid flow occurs. The percolation threshold, the grid blocks in the backbone can be divided to two groups: (i) those in the blobs that are multiply connected and make flow paths very tortuous, and (ii) those that would split the backbone into two parts, if removed, that are called red grid blocks (Pike and Stanley, 1981).

At the core scale, Katz and Thompson (1986) argued that the effective permeability is controlled by the critical pore-throat radius corresponding to the mode of the probability density function of pore throats. Analogously, one may postulate that critical permeability corresponding to the mode of permeability distribution should represent the effective permeability ($k_{\text{eff}}$) of a formation. By comparison with numerical simulations at the reservoir scale, we demonstrate that the mode of permeability distribution accurately estimates the effective permeability in geologic formations with different levels of heterogeneity.

## 2.3. Renormalization group theory

Renormalization group theory (RGT) is another upscaling technique from statistical physics (Reynolds et al., 1977; Stinchcombe and Watson, 1976). Using the analogy between fluid flow through a porous medium and flow of current through an electric circuit, King (1989) mapped a block of cells of different permeabilities into an equivalent resistor network and ultimately to a single resistor. Using this terminology, the effective permeability of a $2 \times 2$ block of isotropic cells was obtained in two dimensions as follows (King, 1989):

$$k_{\text{eff 2D}} = \frac{4(k_1+k_3)(k_2+k_4)[k_2 k_4(k_1+k_3)+k_1 k_3(k_2+k_4)]}{[k_2 k_4(k_1+k_3)+k_1 k_3(k_2+k_4)][k_1+k_2+k_3+k_4]+3(k_1+k_2)(k_3+k_4)(k_1+k_3)(k_2+k_4)} \quad (4)$$

where $k_1, k_2, k_3, k_4$ are permeability values of neighbouring cells used in the 2D renormalization.

In three dimensions, the process of renormalization is more complicated. The fundamental structure is now a 2 × 2 × 2 cube, with uniform pressure on two parallel faces and no flow boundary conditions on the remaining four faces. Several transformations should be performed in order to obtain an equivalent resistance. Green and Patterson (2007) used the idea of splitting a 2 × 2 × 2 cube into four components, treated each component as a two-dimensional block and calculated the effective permeability as follows:

$$k_{\text{eff 3D}}(k_1, k_2, k_3, k_4, k_5, k_6, k_7, k_8) = \frac{1}{4}[k_{\text{eff 2D}}(k_1, k_2, k_3, k_4) + k_{\text{eff 2D}}(k_5, k_6, k_7, k_8) + k_{\text{eff 2D}}(k_5, k_6, k_1, k_2) + k_{\text{eff 2D}}(k_7, k_8, k_3, k_4)] \quad (5)$$

Eq. (5) is only an approximation, and the exact solution is much more complicated (King, 1989).

### 2.4. Effective medium approximation

In the effective medium approximation (EMA), developed by Kirkpatrick (1973), a heterogeneous formation is replaced by a homogeneous one of permeability $k_{\text{eff}}$, which is the same as the permeability of the actual heterogeneous formation. The spatial dependence of permeability in the heterogeneous formation results in local perturbations about the effective permeability of the homogeneous formation. The effective permeability can then be calculated by setting the average perturbation to be zero (Kirkpatrick, 1973)

$$\int \frac{k - k_{\text{eff}}}{k + \left(\frac{z}{2} - 1\right)k_{\text{eff}}} f(k) dk = 0 \quad (6)$$

where *f(k)* is the probability density function of permeability, and *z* is the coordination number equal to 4 and 6 respectively in two and three dimensions. We should note that Eq. (6) with *z* = 4 in two dimensions and 6 in three dimensions reduces to the self-consistent approximation. However, the identification of the coordination number with dimension is more complex. For non-Cartesian grids, one should expect different results.

## 3. Objectives

Critical path analysis (CPA) is an upscaling technique from statistical physics. Although CPA has been widely applied to estimate the $k_{eff}$ at the core (or discrete) scale (Katz and Thompson, 1986; Ghanbarian et al., 2016; Ghanbarian, 2020), to the best of the authors' knowledge its applications at the reservoir (or continuum) scale are very limited (Hunt and Idriss, 2009; Shah and Yortsos, 1996). Therefore, the objectives of this study are to: (1) develop a novel approach for applying the concept of CPA to the estimation of effective permeability at the reservoir scale, (2) evaluate this CPA approach by comparing the effective permeability estimated by the approach with that determined from numerical simulations, and (3) compare the accuracy of the CPA estimations to that of other theoretic models, such as perturbation theory, effective-medium approximation, and renormalization group theory. To achieve our objectives, we focus this research on a wide range of aquifers/reservoirs with different levels of heterogeneity.

## 4. Methodology

### 4.1. Heterogeneity due to spatial variation in permeability

In geological formations and reservoirs, permeability varies spatially. Measurements on cores sampled at different points in geologic formations from various studies indicate that permeability measurements should approximately conform to the log-normal distribution (Haneberg, 2012; Wainwright and Mulligan, 2013)

$$f(k) = \frac{A}{\sqrt{2\pi}\sigma k} exp\left[-\left(\frac{\ln(\frac{k}{k_g})}{\sqrt{2}\sigma}\right)^2\right], \quad k_{min} \leq k \leq k_{max} \tag{7}$$

where $A$ is a normalizing factor, $\sigma$ is the standard deviation of log-permeability, $k_g$ is the geometric mean, and $k_{min}$ and $k_{max}$ are the minimum and maximum permeability values in the formation, respectively. According to Fogg (2010), $\sigma^2$ value can be as large as 10 to 15 in natural geological formations. Moreno and Tsang (1994) also numerically studied effective permeability in media with $\sigma$ as large as 6.

In Fig. 1, we show Eq. (7) and its fit to permeability measurements from the Borden site (Sudicky, 1986). As can be seen, the log-normal distribution with $k_g = 1.5 \times 10^{-11} \text{m}^2$, $\sigma = 0.56$, $k_{min} = 6.1 \times 10^{-14}$ m², and $k_{max} = 3.2 \times 10^{-11}$ m² characterizes the permeability histogram reasonably well with $R^2 = 0.80$.

Based on the results shown in Fig. 1, we generated nine other formations using the same truncated log-normal distribution but different values of $k_g$, $\sigma$, $k_{min}$, and $k_{max}$ as reported in Table 1. Formation 1 is based on the actual measurements from Sudicky (1986) with $k_g = 1.5 \times 10^{-11} m^2$ and $\sigma = 0.56$ presented in Fig. 1. Formation 2 is similar to Formation 1, however, its $k_g$ value is one order of magnitude smaller. All other formations were designed so that a wide range of $k_g$, $\sigma$, $k_{min}$, and $k_{max}$ values can be examined. By designed, we mean that we generated random fields on a domain with the given statistics. As can be deduced from Table 1, $k_g$ and $\sigma$ values span nearly 21 and 2 orders of magnitude, respectively, covering a wide range of formations with various levels of heterogeneity. Furthermore, the value of permeability ranges between $6.1 \times 10^{-14}$ and $3.2 \times 10^{-11}$ $m^2$ in Formations 1 to 5 and from $6.1 \times 10^{-14}$ to $3.2 \times 10^{-5}$ $m^2$ in Formations 6 to 10, which indicates broader permeability distributions and, thus, higher levels of heterogeneity and fluctuations in the latter. In Table 1, we also present the value of $\sigma_Y$ and $(lnk)_{ave}$, the standard deviation and mean of the Normal distribution fitted to the natural logarithm of permeability values ($lnk$) in each formation. As expected, $\sigma$ and $\sigma_Y$ are identical (Table 1). In Formations 1 to 5, the natural logarithm of $k_g$ is very close to $(lnk)_{ave}$. However, the effect of truncation in the permeability distribution is more profound in Formations 6 to 10 causing differences between the natural logarithm of $k_g$ and $(lnk)_{ave}$ values. The values $\sigma_Y = 0.56$ and $(lnk)_{ave} = -25.24$ reported for Formation 1 in Table 1 are very close to $\sigma_Y = 0.585$ and $(lnk)_{ave} = -25.67$ found by Turcke and Kueper (1996) who also fitted the log-normal distribution to experimental data from the Borden site.

**4.2. Numerical simulations in uncorrelated geologic formations**

COMSOL provides a powerful computational finite element-based platform for simulations of flow and transport. The Multiphysics package of COMSOL is capable of generating both two- and three-dimensional geometries on which the simulations can be performed. Fig. 2 shows a 3D domain composed of cells of the same size, where the number of cells along each side of the domain represents the domain size. For example, Fig. 2 (left plan) shows a domain of size 20.

It is well documented in the literature that numerical simulations are scale-dependent (Sahimi, 2011), which means that the numerically simulated permeability is expected to vary with the domain size. Accordingly, the fluid flow simulations need to be carried out at various domain sizes to find the representative elementary volume (REV), the smallest domain size above which the effective permeability does not vary with size. There exist two approaches to study the scale dependence of permeability: (1) fixing the domain size and decreasing the cell size, or (2) increasing the domain size by increasing the cell size. In the former, the number of cells increases while in the latter the number of cells does not vary. In this study, we applied the first approach. For the 2 and 3D simulations of flow, respectively, square and cubic domains of length 10 m were created. We should clarify that the cells refer to the grid blocks used to represent the permeability field. The physical length of 10 m is arbitrary, and any other value can be used without affecting our simulations and results since we determine the REV value of permeability. Each cell in the domain was then randomly assigned a specific value of permeability from the log-normal probability density function. Fig. 2 (right plan) presents the spatial distribution of permeability for the same domain depicted in the same figure. We should point out that our focus is on geologic formations with an uncorrelated distribution of permeability.

Using the finite element analysis, flow was simulated through the 2 and 3D formations by COMSOL, which solves the pressure form of Darcy's law together with the mass conservation equation. To discretize the pressure within the domain, we used the second order of elements. The effect of mesh size on the simulated permeability was investigated based on which the mesh size was selected to be the same as the grid block (normal mesh size in COMSOL). Although finer meshing might be more suitable, we found that smaller mesh sizes resulted in highly computationally demanding simulations (results not shown). For all the simulations, hydraulic head boundary conditions were applied along the flow direction with no-flow conditions applied in the perpendicular directions. The hydraulic head was set equal to 2 m at one side of the flow direction and to 0 at the other side to obtain the hydraulic gradient of 0.2 (Cullen et al., 2010; San and Rowe, 1993). The dynamic viscosity and the water density were set equal to $8.9 \times 10^{-4}\ Pa.s$ and $1000\ kg/m^3$, respectively.

To find the REV, different domain sizes were used. At each domain size, the effective permeability was computed by running simulations 60 times and then averaging over all the iterations to remove the bias in the simulations. The effective permeability was plotted against the domain size to determine the $k_{\text{eff}}$ above the REV.

### 4.3. Estimating $k_{eff}$ via theoretical models

In this study, we evaluate several theoretic models for estimating the effective permeability from the permeability distributions for 10 formations with different levels of heterogeneity (Table 1). To estimate the $k_{eff}$ via the perturbative methods i.e., ANPT, Eq. (1), SPT, Eq. (2), and ALPT, Eq. (3), we used the log-normal permeability distribution parameters given in Table 1. For the ANPT model, we set $\chi = 1/3$ and $\gamma_i = 0$ for isotropic formations as described by Sanchez-Vila et al. (2006) and Indelman and Abramovich (1994). For detailed discussion of the assumptions, strengths, and limitations of perturbative-based models, the interested reader is referred to the review paper of Sanchez-Vila et al. (2006).

To estimate the $k_{eff}$ using the CPA, we determined the value of effective permeability corresponding to the mode of the log-normal permeability distribution using the expression $k_{eff} = \exp[\ln(k_g) - \sigma^2]$.

For the RGT, we constructed two- and three-dimensional matrices in MATLAB whose elements were randomly selected from the log-normal permeability distribution. The dimensions of such matrices were determined based on the REVs. To compute the effective permeability in two and three dimensions, permeability was scaled up at the $2 \times 2$ block and $2 \times 2 \times 2$ cube levels using Eqs. (4) and (5), respectively. For each geologic formation, we iterated these computations 1000 times and averaged over all to calculate the $k_{eff}$.

To estimate the effective permeability within the EMA framework, we numerically solved Eq. (6) in MATLAB. For this purpose, we used the trapezoidal numerical integration, which approximately computes an integral via the trapezoidal method with unit spacing. By trial and error, we found that the integral in Eq. (6) could be well approximated using 1000 trapezoids. In two and three dimensions, we set $z = 4$ and 6, respectively. All the MATLAB codes used in this study can be found in Adeyemi (2021).

### 4.4. Models evaluation criteria

To evaluate the accuracy of each model, we used the root mean square log-transformed error (RMSLE) because the estimated effective permeability may vary over a wide range. We also computed the relative error (RE) values. The two statistical parameters were calculated as follows

$$RMSLE = \sqrt{\frac{1}{N} \sum_{i=1}^{N} [\log_e(x_{\text{est}}) - \log_e(x_{\text{sim}})]^2} \tag{8}$$

$$RE = \frac{x_{\text{est}} - x_{\text{sim}}}{x_{\text{sim}}} \times 100 \tag{9}$$

where $N$ is the number of samples, $x_{\text{est}}$ and $x_{\text{sim}}$ are, respectively, the estimated and simulated values, and $\log_e$ represents the natural logarithm.

### 5. Results

In this section, we compare the estimated effective permeability values by different models including the ANPT, Eq. (1), SPT, Eq. (2), ALPT, Eq. (3), CPA, RGT, Eqs. (4) and (5), and EMA, Eq. (6), with the numerically simulated values from COMSOL in Figs. 3 and 4 for the two- and three- dimensional geologic formations, respectively. In Formations 1 through 5, permeability spans about three orders of magnitude ($6.1 \times 10^{-14} \leq k \leq 3.2 \times 10^{-11}$), and $\sigma$ ranges from 0.05 and 0.56. In Formations 6 to 10, however, permeability spans nearly eight orders of magnitude ($6.1 \times 10^{-14} \leq k \leq 3.2 \times 10^{-5}$), and $\sigma$ varies between 2 and 6 (Table 1). Formations 1 to 5 represent relatively heterogeneous reservoirs, while Formations 6 to 10 denote heterogeneous systems.

In our study, Formation 1 statistically represents the isotropic version of the Borden site. Using the horizontal and vertical permeability values of $0.84 \times 10^{-11}$ m$^2$ ($8.2 \times 10^{-3}$ cm/s) and $0.65 \times 10^{-11}$ m$^2$ ($6.33 \times 10^{-3}$ cm/s), respectively, Sudicky (1986) found that the anisotropy ratio was about 1.3 in the Borden site. Interestingly, the two- and three-dimensional simulated effective permeability values presented in Table 1 are only 25 and 34% greater than the horizontal permeability reported by Sudicky (1986). This shows reasonable agreement between our numerical simulations and the experimental measurements.

In what follows, we address the reliability and accuracy of each model based on its performance in this study. The REV plots, based on which the representative permeability value for each formation was determined, are presented in Figs. (1A) and (2A) in Appendix A for both two and three dimensions.

### 5.1. Perturbative methods

In two dimensions, all of the perturbative models i.e., ANPT, SPT, and ALPT estimated the effective permeability accurately in Formations 1 to 5 (with $\sigma \leq 0.56$). However, they substantially overestimated the $k_{\text{eff}}$ in Formations 6 to 10 (with $\sigma \geq 2$) as shown in Figs. 3a-3c. The RE values of the $k_{\text{eff}}$ estimations by each model are reported in Table 2. We found RMSLE = 15.32, 15.32, and 5.49 respectively for the ANPT, Eq. (1), SPT, Eq. (2), and ALPT, Eq. (3), models. We should note that, in two dimensions, both the ANPT, Eq. (1), with $\gamma_i = 0$ and $\epsilon = 0$, and SPT, Eq. (2), reduce to Matheron's conjecture (Matheron, 1967) in which $k_{\text{eff}} = k_{\text{g}}$. As a result, the ANPT and SPT models resulted in the same estimations with RMSLE and average RE values of 15.32 and 1.72×10$^{17}$% for all the formations. We also investigated models' accuracy within Formations 1 to 5 and 6 to 10. For the ANPT model, we found RMSLE = 0.22 and 21.66. Same results were obtained for the SPT model. For the ALPT model RMSLE = 0.22 and 7.76 were found respectively for Formations 1 to 5 and 6 to 10.

Similar results were obtained in three dimensions; the three perturbative methods overestimated the effective permeability in Formations 6 to 10, while they provided accurate estimations in Formations 1 to 5 (Figs. 4a-4c). Table 3 lists the RE values for the estimated $k_{eff}$ values by each model. We found RMSLE values of 15.77, 14.30, and 4.21 for the ANPT, Eq. (1), SPT, Eq. (2), and ALPT, Eq. (3), models respectively. While the performance of the ANPT model in three dimensions deteriorated compared to its performance in two dimensions, the SPT model performed slightly better. The accuracy of the ALPT also improved from two to three dimensions (RMSLE = 5.49 and 4.21 respectively).

We also compared the performance of the perturbative methods within Formations 1 to 5 and 6 to 10. Comparison of the ANPT estimations with COMSOL simulations in Formations 1 to 5 and 6 to 10 showed that this model reasonably estimated $k_{eff}$ in the former (relatively heterogeneous formations) with RMSLE of 0.22, while it overestimated the $k_{eff}$ with RMSLE = 20.18 in the latter (heterogeneous formations).

For the SPT model, we found RMSLE = 0.18 and 20.18 and average RE values = 16 and $5 \times 10^{16}$ % in Formations 1 to 5 and 6 to 10, respectively. For the ALPT model, however, RMSLE = 0.22 and 5.94, values were less, particularly in Formations 6 to 10, compared to the SPT and ANPT models. The ALPT model underestimated $k_{eff}$ with an average RE of 19.9% in Formations 1 to 5 and overestimated the effective permeability with an average RE of $2.93 \times 10^6$%, about ten orders of magnitude smaller than that obtained from the SPT model in Formations 6 to 10. Our results demonstrated that the ALPT model estimates the effective permeability accurately in three-dimensional formations with $\sigma \leq 4$.

**5.2. Critical path analysis**

Two-dimensional results from the CPA are presented in Fig. 3d. As can be seen, the CPA with RMSLE = 0.50 estimated the $k_{eff}$ in all formations accurately (with data points around the 1:1 line indicating good agreement between the numerical simulations and the CPA estimations). We found an average RE value of 19.8% for all formations. Although in most formations the CPA slightly underestimated the effective permeability, it overestimated $k_{eff}$ in Formation 10 with RE = 299.21% (Table 2). Further analysis showed that the CPA estimated $k_{eff}$ with RMSLE = 0.02 and 0.70 within Formations 1 to 5 and 6 to 10, with average RE values of 0.4% and 39.21%, respectively. It is worth pointing out that the RE values calculated from the $k_{eff}$ estimations for the relatively heterogenous formations are all less than 5%, with the highest relative error of 4.41% in Formation 1 (Table 2).

In three dimensions, CPA also estimated $k_{eff}$ accurately with RMSLE = 1.14 and average RE = -36.8%. Fig. 4d shows the estimated effective permeability values against the simulated ones in three dimensions. We should point out that the CPA estimations in two and three dimensions are the same since the mode of the permeability distribution does not vary with formation dimensionality. Generally speaking, the CPA tended to underestimate the $k_{eff}$ in most 3D formations (Table 3). The CPA estimated $k_{eff}$ within Formations 1 to 5 with RMSLE = 0.04 and average RE = -2.61%. Within Formations 6 to 10, the values of RMSLE and average RE were 1.61 and -70.94%, respectively.

### 5.3. Renormalization group theory

The RGT estimated the effective permeability in 2D formations accurately with RMSLE = 0.67 (Fig. 3e) and average RE of -27%. It can be deduced from the average RE value that the RGT model, Eq. (5), on average, underestimated the $k_{eff}$. Similar to the CPA, the RGT estimated $k_{eff}$ in Formations 1 to 5 more precisely with RMSLE = 0.012 and average RE = -0.9%. These values, however, were 0.94 and -53% in Formations 6 to 10.

In three dimensions, the RGT estimated the effective permeability in all ten formations with RMSLE = 0.90 (Fig. 4e) and average RE of -35% (slightly more accurate than the CPA). Comparison of the estimated $k_{eff}$ values with the simulated ones resulted in RMSLE = 0.03 and 1.27, and average RE = -3% and -67% in Formations 1 to 5 and 6 to 10, respectively. The lowest and highest relative error belong respectively to Formations 5 and 9 (Table 3). Similar to the results of CPA, the RGT model tended to underestimate the effective permeability in most formations studied here.

### 5.4. Effective medium approximation

The $k_{eff}$ estimations via the EMA against the $k_{eff}$ simulations by COMSOL are shown in Fig. 3f for two-dimensional formations. We found RMSLE = 5.03 and average RE = $2.81 \times 10^5$%. As can be seen from Fig. 3f, the EMA accurately estimated the effective permeability in Formations 1 to 5 with RMSLE = 0.20 (similar to the perturbative methods) and average RE = 18%. This can be visually confirmed from the data points that lay on the 1:1 line in Fig. 3f. Although the EMA underestimated $k_{eff}$ in Formation 6 ($\sigma = 2$), it overestimated the effective permeability in Formations 7 to 10 with $\sigma \geq 3$. For Formations 6 to 10, however, we found RMSLE = 7.11 with an average relative error value of $5.6 \times 10^5$ %.

Results of the $k_{\text{eff}}$ estimations by the EMA and the numerically simulated $k_{\text{eff}}$ values by COMSOL in three dimensions are presented in Fig. 4f. We found RMSLE = 4.27 and average RE = $7.8 \times 10^4$% for all formations. The EMA estimated the $k_{\text{eff}}$ in Formations 1 to 5 accurately. However, it underestimated the effective permeability in Formation 6 and overestimated that in Formations 7 to 10. Comparing the RMSLE values from 2 and 3D results (5.03 vs. 4.27 respectively) shows that the EMA provided more accurate estimations in three dimensions. Further comparison rendered RMSLE = 0.19 in Formations 1 to 5 (similar to the perturbative methods) and RMSLE = 6.03 in Formations 6 to 10. The average relative errors were 17% and $1.6 \times 10^5$% in Formations 1 to 5 and Formations 6 to 10 respectively.

## 6. Discussion

### 6.1. Models performance

King (1989), Renard and de Marsily (1997), Sanchez-Vila et al. (2006) and many others pointed out that perturbative methods provide accurate estimations of the effective permeability only in media with small variations in permeability. More specifically, results by Hristopulos and Christakos (1999) and Dykaar and Kitanidis (1992) showed that Matheron's conjecture and other perturbative methods can be successfully applied to formations with $\sigma = 2.65$ and smaller. Similarly, evidence from this study showed substantial effective permeability overestimations by several perturbative methods in formations with $\sigma \geq 2$.

Many perturbative methods reduce to the exact form of Matheron's conjecture ($k_{\text{eff}} = k_g$) in two dimensions. Although the ANPT and SPT models include terms in their expressions other than the permeability standard deviation, these models reduce to $k_{\text{eff}} = k_g$ when applied to two-dimensional isotropic geologic formations. Therefore, it is not surprising that the perturbative methods used in this study largely overestimated the effective permeability in the heterogenous geologic formations with $\sigma \geq 2$. However, the ALPT model does not reduce to the exact form of the conjecture, and this, in addition to the inverse form of its $k_{\text{eff}}$ expression, is likely the reason for the higher accuracy of this model.

The CPA approach has been successfully applied to estimate the $k_{\text{eff}}$ at the core scale (Katz and Thompson, 1986; Ghanbarian et al., 2017; Ghanbarian, 2020). However, its applications at the field scale are very limited in the literature. Hunt and Idriss (2009) applied concepts from CPA to determine the effective permeability in correlated and random systems with bimodal permeability distributions in terms of the arithmetic mean of $k_{min} < k < k_{max}$ and harmonic mean of $k_c < k < k_{max}$ in which $k_c$ is the critical permeability (see their Eq. 8). They showed that the CPA provided reasonable estimations above the percolation threshold in correlated systems.

Recently, Masihi et al. (2016) determined the connected cluster of cells with high permeabilities in a permeability field to calculate a threshold value. For this purpose, they started with a high permeability value from its distribution and evaluated whether the cells with permeabilities greater than the selected permeability resulted in a connected path along their domain. If not, a smaller permeability value was selected, and the process was repeated until the sample-spanning cluster formed. The permeability value corresponded to the presence of the first percolating cluster was called the threshold permeability and compared with numerical simulations of the effective permeability. Masihi et al. (2016) reported that the threshold permeability did not estimate the effective permeability accurately (see their Fig. 8). They also found that the threshold permeability was less than the permeability corresponded to the mode of permeability distribution (their Fig. 2). Further investigations are still required to determine the threshold permeability and its relationship with the effective permeability for a wide range of formations with different levels of heterogeneities.

Several studies in the literature have highlighted the reliability of the RGT for estimating the effective permeability in homogenous and heterogeneous formations (King, 1989; Hristopulos and Christakos, 1999; Green and Paterson, 2007). Nonetheless, real space renormalization does have some systematic errors. This is partly due to the small cell representation (one could use larger cells than 2×2 or use a different finite difference representation or boundary conditions for the small cell). Some of these errors may eliminate in three dimensions because of the greater freedom for flow, and the less impact by boundary conditions, as our results demonstrated. It indeed depends a bit on the permeability distribution and, in particular, the correlation structure. One may find larger errors, if there are substantial contrasts between neighboring cells, especially if they are spatially extended.

It is well documented in the literature that the EMA returns accurate estimations for small variances (Adler and Berkowitz, 2000; Ghanbarian and Daigle, 2016). For example, Adler and Berkowitz (2000) evaluated the accuracy of the EMA in the estimation of electrical conductivity in two- and three-dimensional media with local conductances that followed the log-normal distribution of various standard deviations. They concluded that, "... the analytical expressions [the effective-medium approximations] provide good agreement to the simulations in 2D systems, but are in significant error in 3D systems when the standard deviation of the local conductivities is large."

The CPA and RGT models provided the most accurate estimations of $k_{\text{eff}}$ in the two- and three-dimensional formations studied here. Both models precisely estimated the effective permeability in relatively heterogeneous (Formations 1 to 5) and heterogeneous (Formations 6 to

10) reservoirs. While the RGT approach requires coding and computations with realizations, the CPA model simply estimates the effective permeability from the mode of the permeability distribution, which saves time and computations.

### 6.2. 2D versus 3D simulations

Results showed that the $k_{\text{eff}}$ values from two- and three-dimensional simulations were highly correlated, with the relation $(k_{eff})_{3D} = 234.14(k_{eff})_{2D}^{1.2}$ and $R^2 = 0.99$. In all formations except Formation 5, the value of $k_{\text{eff}}$ in three dimensions was greater than that in two dimensions (Table 1). This is consistent with the results of King (1989) and Adler and Berkowitz (2000). More specifically, King (1989) simulated the $k_{\text{eff}}$ in two- and three-dimensional systems with uniform and log-normal permeability distributions and reported $k_{\text{eff}}$ in three dimensions to be greater than that in two dimensions.

### 6.3. Long-range correlation and anisotropy

It is well documented in the literature that there might exist long-range correlation at the aquifer/reservoir scale (Clark et al., 2020; Sahimi, 2011; Sahimi and Mukhopadhyay, 1996). Correlation means that heterogeneity e.g., permeability in one zone of a geologic formation is not fully independent of that in other zones. Sahimi (1994) stated that natural porous media are not always random and may exhibit some correlation. For instance, core-scale porous media may contain only short-range correlations, while heterogeneous field-scale systems, such as aquifers and reservoirs, may be long-range correlated. By comparing simulations in correlated and random formations, Hunt and Idriss (2009) found effective permeability in correlated media greater than that in similar uncorrelated media. This is because the value of percolation threshold in correlated formations is smaller than that in random media (Hunt and Idriss, 2009). We should point out that if there is finite-range correlation and the domain is much larger than the correlation length there should be no change in the percolation threshold. However, if the domain is smaller than or comparable to the correlation length, then there should be a shift in the apparent threshold (Masihi and King, 2007a, 2007b). If the correlation is not finite range (e.g., power law) then there is a consistent change in the percolation threshold (King and Masihi, 2018).

Geologic formations might also be anisotropic at such scales. In general, there might be three types of anisotropy in geological structures: (I) anisotropy due to presence of randomly oriented anisotropic permeability blocks, (II) anisotropy due to a direction-dependent permeability distribution, and (III) anisotropy due to the presence of permeability zones of different orientations with different probabilities of flow availability (Mukhopadhyay and Sahimi, 2000). In type I, the effective permeability of such formations is always isotropic. In type II, the anisotropy may vanish under certain circumstances, while in type III anisotropy always remains (Mukhopadhyay and Sahimi, 2000).

In this study, we evaluated the CPA approach in isotropic and uncorrelated (random) formations. Further investigations are required to assess the reliability and predictability of the CPA in anisotropic and correlated large-scale porous media.

### 6.4. Unimodal versus multimodal permeability distributions

The proposed CPA-based approach to determine the effective permeability in random geologic formations is based on unimodal permeability distribution. However, permeability field in aquifers or reservoirs may conform to multimodal distributions (Hunt and Idriss, 2009;

Massabó et al., 2008; Rubin, 1995; Tidwell and Wilson, 1999). Extending the proposed CPA to media with bimodal and multimodal permeability distributions is still required.

## 7. Conclusions

Using concepts from critical path analysis (CPA), we presented a novel approach for estimating the effective permeability of a geologic formation. Based on the CPA, low permeability zones in a formation contribute little to fluid flow, while the high permeability zones significantly influence the flow of fluids. Based on this principle, we postulated that permeability at the mode of permeability density function should represent the effective permeability of a reservoir. The proposed CPA approach was evaluated by comparing its effective permeability estimations for the two- and three-dimensional formations in this study with numerically simulated effective permeability values for the formations. The log-normal distribution with different geometric means ($4.5\times10^{-12} \leq k_g \leq 1.0\times10^9$ m$^2$) and standard deviations ($0.05 \leq \sigma \leq 6$) was used to generate such formations. In addition to the CPA, other theoretic approaches, such as perturbative methods, renormalization group theory, and effective medium approximation, were applied to estimate the $k_{\text{eff}}$. Results showed that the CPA estimated the $k_{\text{eff}}$ with RMSLE = 0.50 more accurate than the other approaches in two dimensions. However, the RGT with RMSLE = 0.90 estimated the $k_{\text{eff}}$ slightly more accurately than the CPA with RMSLE = 1.14 in three dimensions. We also found that the perturbative methods and EMA provided reasonable estimations of $k_{\text{eff}}$ in formations with $\sigma \leq 2$. However, these approaches substantially overestimated the effective permeability in highly heterogeneous formations with $\sigma > 2$. The heterogeneity range tested in this work is large, and we find the CPA as a powerful platform to estimate the effective permeability at the reservoir scale in uncorrelated formations. However, further investigations are required to evaluate the reliability of this CPA approach in correlated and anisotropic geologic formations.

## Acknowledgements

BA is grateful to National Association of Black Geoscientists (NABG), Kansas Geological Foundation (KGF), and the Kansas State University Graduate Students Council for scholarships and travel grants. BG acknowledges Kansas State University for supports through faculty startup fund and university small research grant (USRG).

Table 1. Ten different geological formations constructed in this research study. $k_g$ is the geometric mean and $\sigma$ is the standard deviation from the log-normal permeability distribution. $k_{min}$ and $k_{max}$ are the minimum and maximum permeability values in each formation. $(\ln k)_{avg}$ is the average of the natural logarithm of permeability and $\sigma_Y$ is the standard deviation of log-transformed permeability distribution ($Y = \ln(k)$). $k_{eff}$ represents the effective permeability simulated by COMSOL in two and three dimensions.

| Formation | $k_g$ (m²) | $\sigma$ | $k_{min}$–$k_{max}$ | $(\ln k)_{avg}$ | $\sigma_Y$ | $k_{eff}$ (m²) 2D | $k_{eff}$ (m²) 3D |
|---|---|---|---|---|---|---|---|
| 1 | $1.5\times10^{-11}$ | 0.56 | $6.1\times10^{-14}$–$3.2\times10^{-11}$ | -25.24 | 0.56 | $1.05\times10^{-11}$ | $1.12\times10^{-11}$ |
| 2 | $1.5\times10^{-12}$ | 0.56 | $6.1\times10^{-14}$–$3.2\times10^{-11}$ | -27.54 | 0.56 | $1.11\times10^{-12}$ | $1.16\times10^{-12}$ |
| 3 | $1.9\times10^{-11}$ | 0.25 | $6.1\times10^{-14}$–$3.2\times10^{-11}$ | -24.76 | 0.25 | $1.73\times10^{-11}$ | $1.76\times10^{-11}$ |
| 4 | $4.5\times10^{-12}$ | 0.40 | $6.1\times10^{-14}$–$3.2\times10^{-11}$ | -26.29 | 0.40 | $3.92\times10^{-12}$ | $4.02\times10^{-12}$ |
| 5 | $2.6\times10^{-11}$ | 0.05 | $6.1\times10^{-14}$–$3.2\times10^{-11}$ | -24.37 | 0.05 | $2.62\times10^{-11}$ | $2.61\times10^{-11}$ |
| 6 | $5.0\times10^{-11}$ | 2.0 | $6.1\times10^{-14}$–$3.2\times10^{-5}$ | -27.72 | 2.0 | $1.26\times10^{-12}$ | $2.28\times10^{-12}$ |
| 7 | $1.0\times10^{-6}$ | 3.0 | $6.1\times10^{-14}$–$3.2\times10^{-5}$ | -22.81 | 3.0 | $1.67\times10^{-10}$ | $5.13\times10^{-10}$ |
| 8 | $1.0\times10^{-3}$ | 4.0 | $6.1\times10^{-14}$–$3.2\times10^{-5}$ | -22.9 | 4.0 | $2.03\times10^{-10}$ | $1.20\times10^{-09}$ |
| 9 | $1.0\times10^{3}$ | 5.0 | $6.1\times10^{-14}$–$3.2\times10^{-5}$ | -18.10 | 5.0 | $1.46\times10^{-8}$ | $1.09\times10^{-07}$ |
| 10 | $1.0\times10^{9}$ | 6.0 | $6.1\times10^{-14}$–$3.2\times10^{-5}$ | -15.25 | 6.0 | $5.81\times10^{-8}$ | $3.94\times10^{-07}$ |

Table 2. Relative error (%) calculated in the estimation of $k_{eff}$ by each model and for the 2D Formations.

| Formation | $k_g$(m²) | $\sigma$ | ANPT | SPT | ALPT | CPA | RGT | EMA |
|---|---|---|---|---|---|---|---|---|
| 1 | $1.5\times10^{-11}$ | 0.56 | 43.47 | 43.47 | 43.40 | 4.41 | -0.81 | 35.70 |
| 2 | $1.5\times10^{-12}$ | 0.56 | 35.35 | 35.35 | 35.30 | -1.49 | -2.08 | 34.40 |
| 3 | $1.9\times10^{-11}$ | 0.25 | 8.04 | 8.04 | 8.04 | 1.49 | 0.1 | 7.36 |
| 4 | $4.5\times10^{-12}$ | 0.40 | 14.63 | 14.63 | 14.60 | -2.32 | -1.45 | 14.80 |
| 5 | $2.6\times10^{-11}$ | 0.05 | 0.14 | 0.14 | 0.14 | -0.11 | -0.30 | 0.02 |
| 6 | $5.0\times10^{-11}$ | 2.0 | 3858 | 3858 | 437 | -27.50 | -12.88 | -95.2 |
| 7 | $1.0\times10^{-6}$ | 3.0 | $6.00\times10^5$ | $6.00\times10^5$ | 904 | -25.97 | -49.92 | $4.96\times10^5$ |
| 8 | $1.0\times10^{-3}$ | 4.0 | $4.93\times10^8$ | $4.93\times10^8$ | 6510 | -44.51 | -63.48 | $2.23\times10^6$ |
| 9 | $1.0\times10^3$ | 5.0 | $6.83\times10^{12}$ | $6.83\times10^{12}$ | $3.83\times10^{12}$ | -5.20 | -71.71 | $6.80\times10^4$ |
| 10 | $1.0\times10^9$ | 6.0 | $1.72\times10^{18}$ | $1.72\times10^{18}$ | $1.81\times10^8$ | 299.21 | -68.66 | $2.00\times10^4$ |

Table 3. Relative error (%) calculated in the estimation of $k_{eff}$ by each model and for the 3D Formations.

| Formation | $k_g$(m²) | $\sigma$ | ANPT | SPT | ALPT | CPA | RGT | EMA |
|---|---|---|---|---|---|---|---|---|
| 1 | $1.5\times10^{-11}$ | 0.56 | 41.6 | 34 | 41.40 | -2.18 | 4.72 | 26.30 |
| 2 | $1.5\times10^{-12}$ | 0.56 | 36.2 | 29 | 35.90 | -5.94 | 4.54 | 36.50 |
| 3 | $1.9\times10^{-11}$ | 0.25 | 7.3 | 6 | 7.30 | -0.24 | 1.07 | 58.80 |
| 4 | $4.5\times10^{-12}$ | 0.40 | 14.8 | 12 | 14.80 | -4.76 | 2.675 | 15.00 |
| 5 | $2.6\times10^{-11}$ | 0.05 | 0.37 | 0.33 | 0.37 | 0.08 | 0.06 | 0.21 |
| 6 | $5.0\times10^{-11}$ | 2.0 | 2090 | 2095 | 57.20 | -59.80 | 36.05 | -97.30 |
| 7 | $1.0\times10^{-6}$ | 3.0 | $1.95\times10^5$ | $1.95\times10^5$ | 66.50 | -75.96 | 63.54 | $2.66\times10^5$ |
| 8 | $1.0\times10^{-3}$ | 4.0 | $8.37\times10^7$ | $8.37\times10^7$ | 495.00 | -90.58 | 78.17 | $5.03\times10^5$ |
| 9 | $1.0\times10^3$ | 5.0 | $9.13\times10^{11}$ | $9.13\times10^{11}$ | $2.75\times10^4$ | -87.32 | 80.03 | $1.03\times10^4$ |
| 10 | $1.0\times10^9$ | 6.0 | $2.54\times10^{17}$ | $2.54\times10^{17}$ | $1.46\times10^7$ | -41.05 | 74.80 | $3.20\times10^3$ |

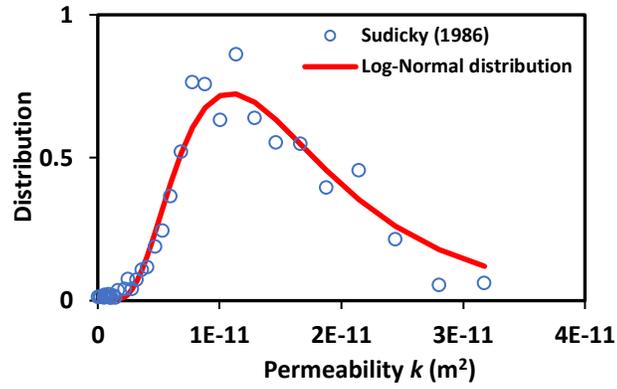

Fig. 1. The log-normal distribution, Eq. (8), with $k_g = 1.5 \times 10^{-11}$ m², $\sigma = 0.56$, $k_{min} = 6.1 \times 10^{-14}$ m², and $k_{max} = 3.2 \times 10^{-11}$ m² fitted with $R^2 = 0.80$ to the permeability histogram. Permeability measurements are from the Borden site (Sudicky, 1986).

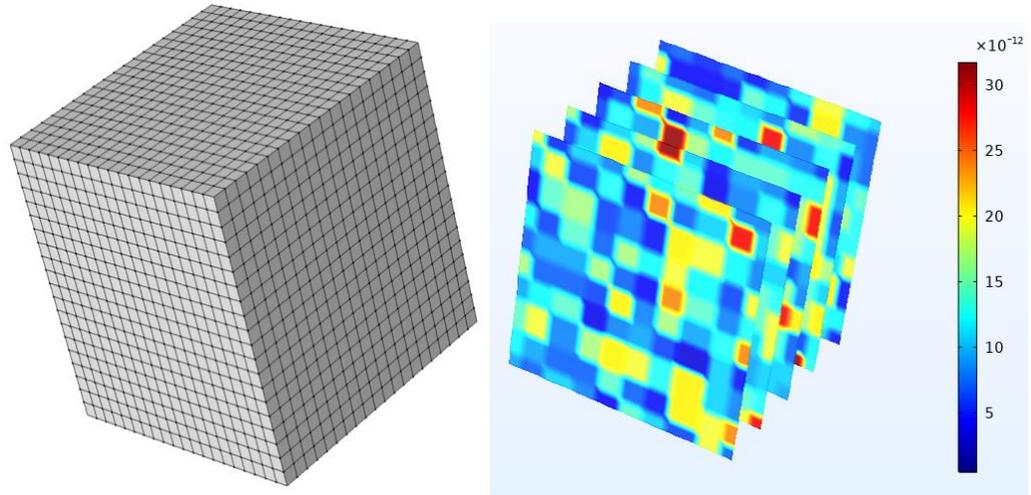

Fig. 2. (left) A 3D domain of size of 10 m with 20 cells along each side (domain size = 20), and (right) random spatial distribution of permeability values in the same domain.

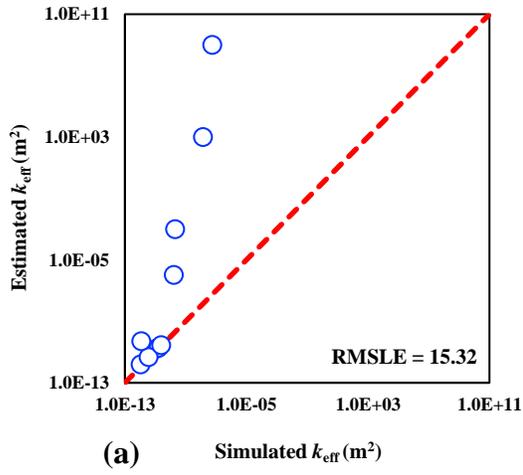 (a)
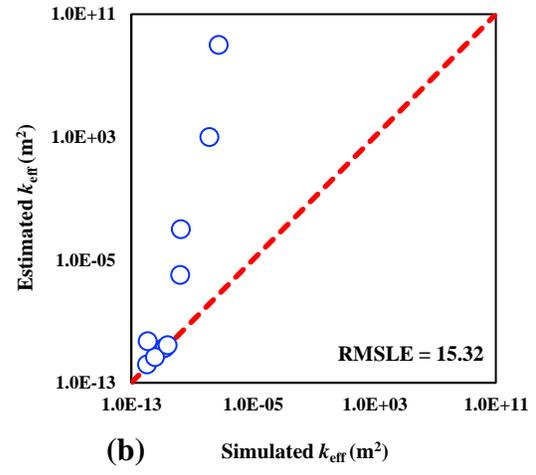 (b)
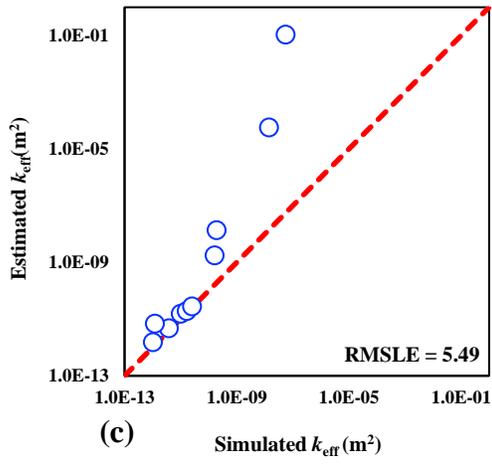 (c)
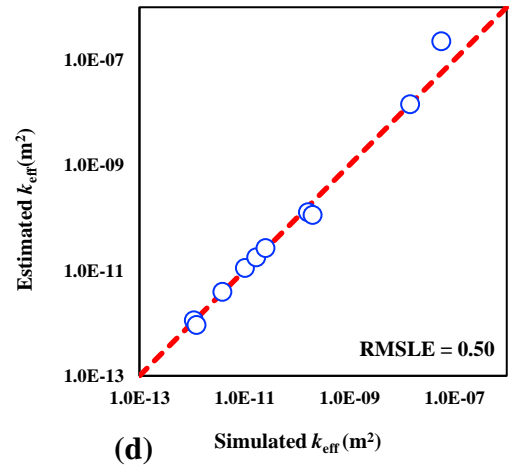 (d)
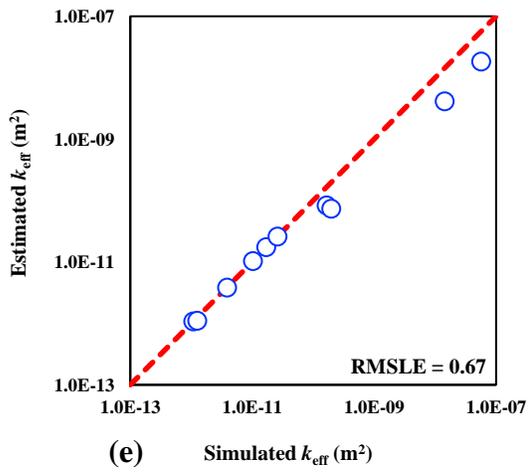 (e)
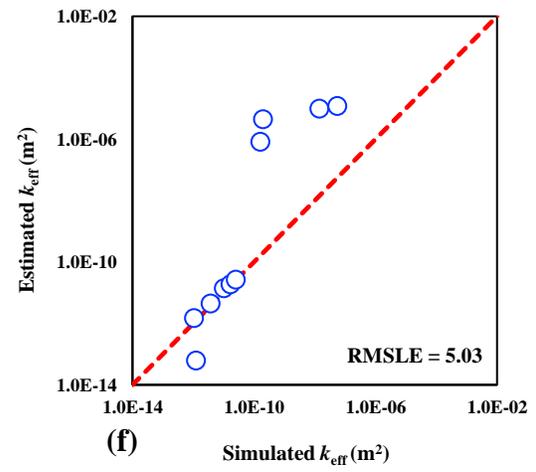 (f)

Fig. 3. Comparison of effective permeabilities calculated from 2D numerical simulations and those estimated from models including (a) ANPT, Eq. (1), (b) SPT, Eq. (2), (c) ALPT, Eq. (3), (d) CPA, (e) RGT, Eq. (4), and (f) EMA, Eq. (6)..

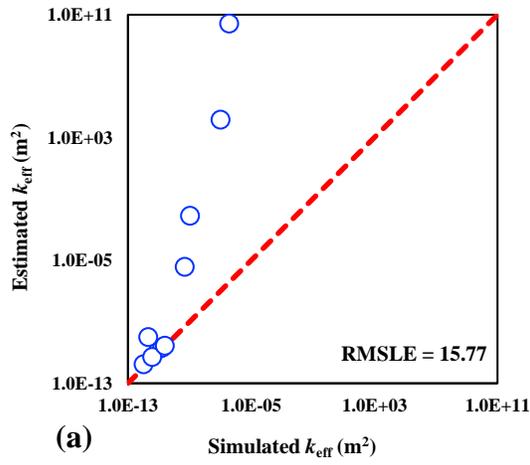
(a)

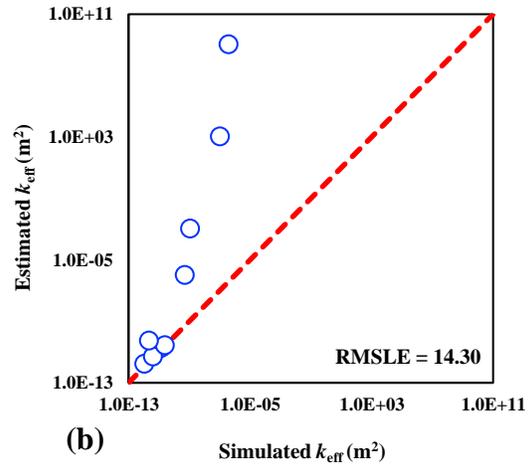
(b)

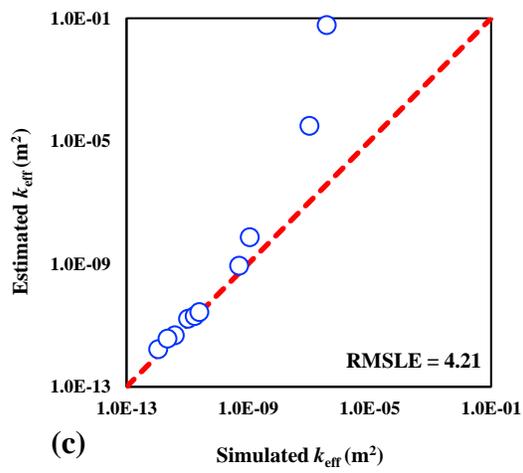
(c)

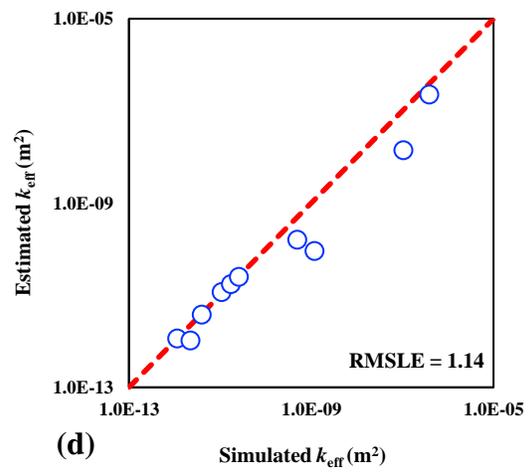
(d)

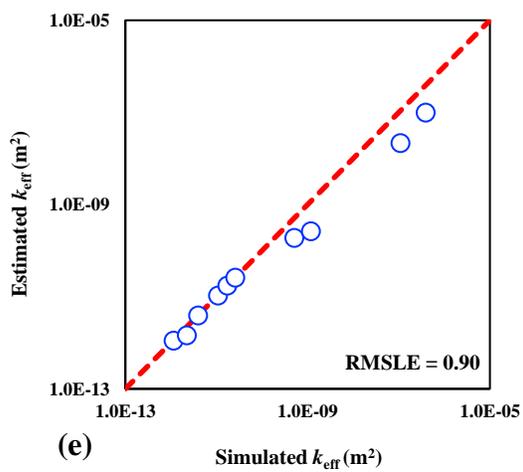
(e)

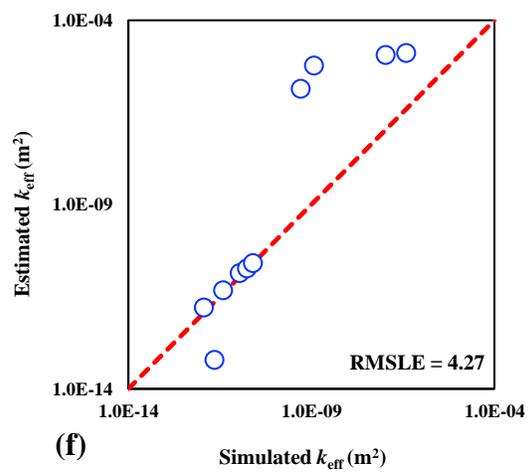
(f)

Fig. 4. Comparison of effective permeabilities calculated from 3D numerical simulations and those estimated from models including (a) ANPT, Eq. (1), (b) SPT, Eq. (2), (c) ALPT, Eq. (3), (d) CPA, (e) RGT, Eq. (4), and (f) EMA, Eq. (6).

# Appendix A

In this appendix, the effective permeability numerically computed via COMSOL against the domain size are presented for the two- and three-dimensional simulations. The domain size indicates the number of cells along each side of domain. The representative elementary volume (REV) was accordingly determined for each formation based on these plots.

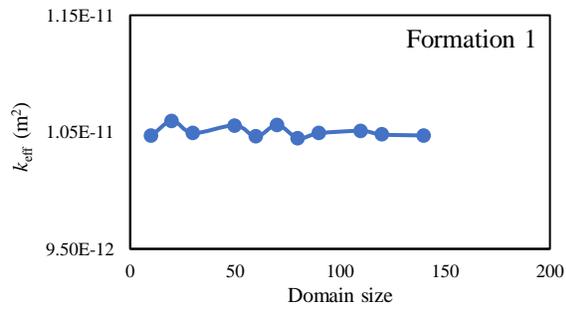
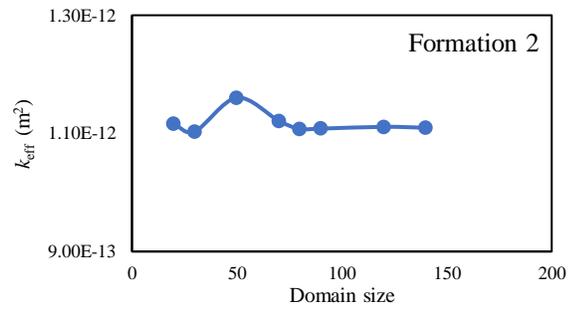
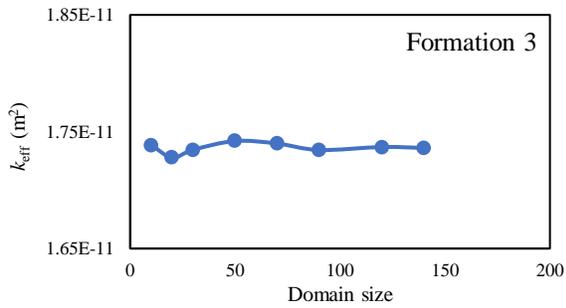
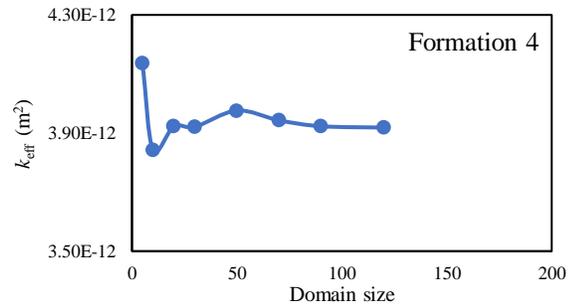
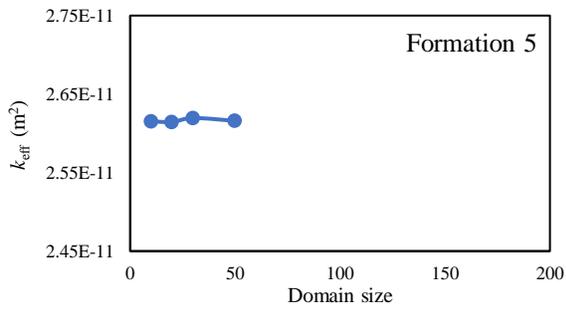
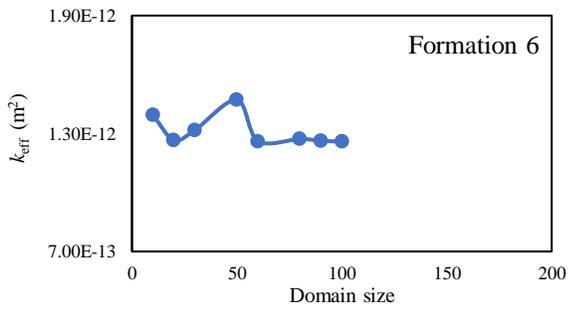

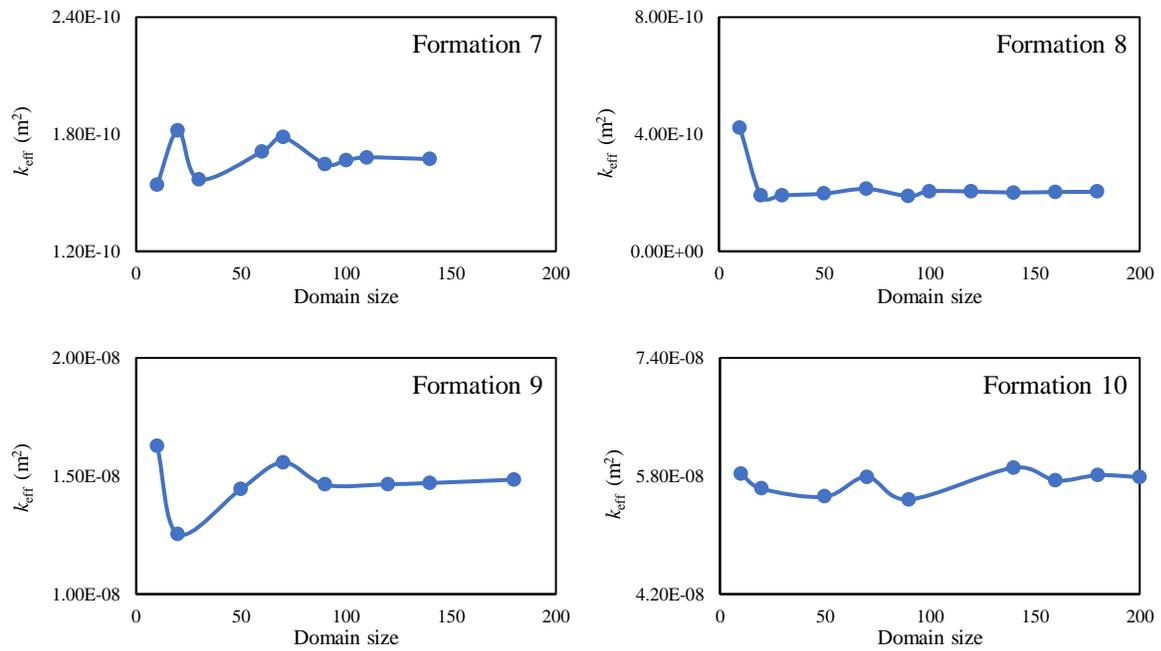

Fig. 1A. Plots of effective permeability against domain size to determine the representative elementary volume (REV) for each of the 2D formations.

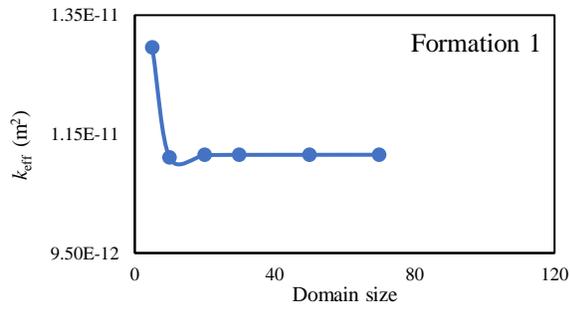
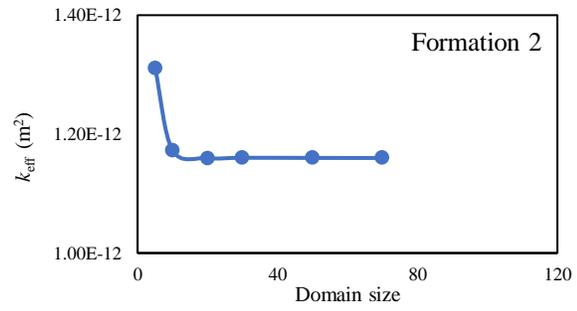
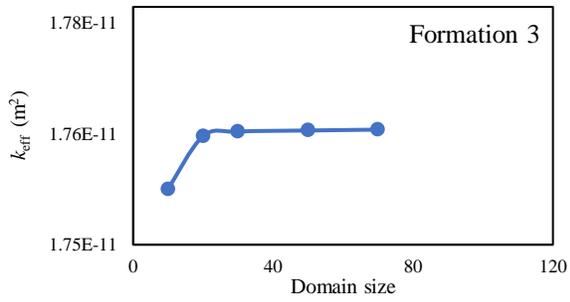
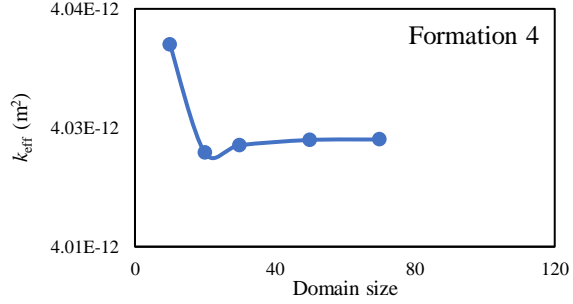
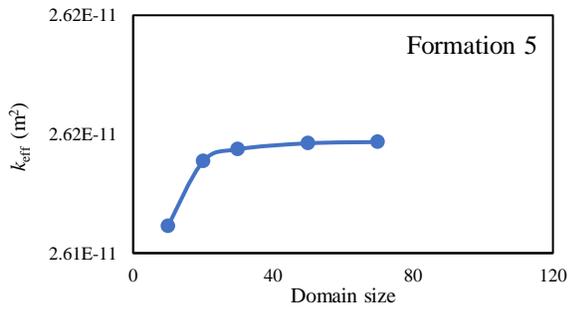
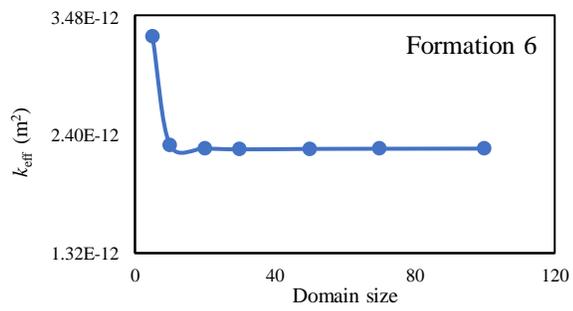
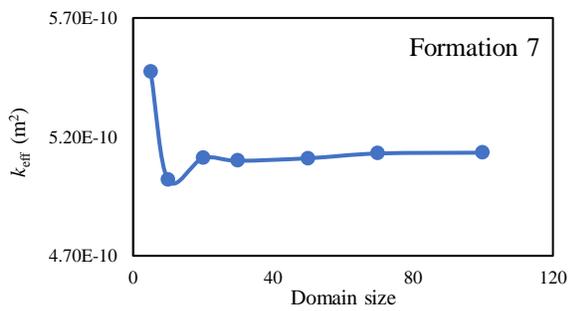
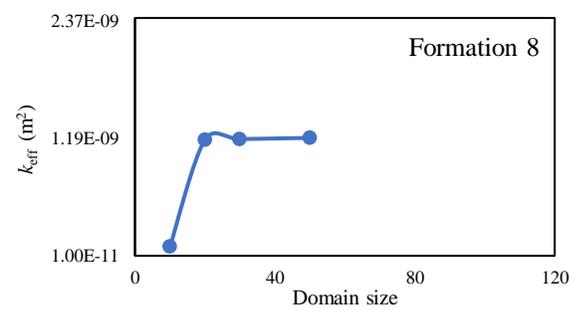
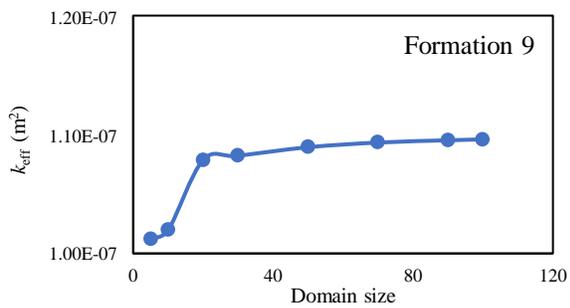
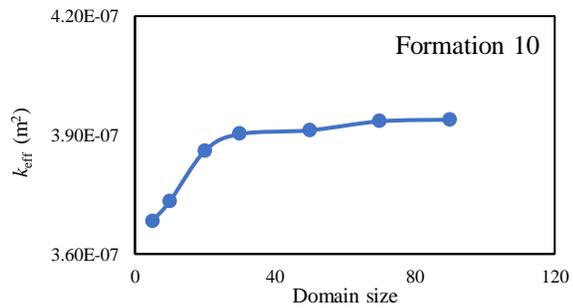

Fig. 2A. Plots of effective permeability against domain size to determine the representative elementary volume (REV) for each of the 3D formations.